\documentclass[12pt]{iopart}
\usepackage{graphicx}
\usepackage{amssymb}
\usepackage{amsthm}
\usepackage{braket}
\usepackage{bbold}
\usepackage{csquotes}
\usepackage[disable]{todonotes}
\usepackage{notes2bib}
\makeatletter
\newcommand{\overset}[2]{\ensuremath{\mathop{\kern\z@\mbox{#2}}\limits^{\mbox{\scriptsize #1}}}}
\makeatother
\def\eqref{\eref}
\def\text{\mathrm}
\newenvironment{align}{\begin{eqnarray}}{\end{eqnarray}}
\newenvironment{multline}{\begin{eqnarray}}{\end{eqnarray}}

\newcommand{\eg}{e.g.\@}
\newcommand{\cf}{c.f.\@}
\newcommand{\ie}{i.e.\@}
\newcommand{\pseudoP}{q}
\newcommand{\erf}{\mathop{\mathrm{erf}}}
\newcommand{\f}{f_{0}}
\newcommand{\intSp}[1][i]{\bar{S}_{#1}^{+}}
\newcommand{\opSp}[1][i]{\hat{S}_{#1}^{+}}
\newcommand{\intSm}[1][i]{\bar{S}_{#1}^{-}}
\newcommand{\opSm}[1][i]{\hat{S}_{#1}^{-}}
\newcommand{\intr}{I_{r}}
\newcommand{\intl}{I_{l}}
\newcommand{\xseam}{x_{\mathrm{s}}}
\newcommand{\kb}{k_{\mathrm{B}}}
\newcommand{\xinit}{x_{\mathrm{init}} }

\begin{document}

\title{Propagator for a driven Brownian particle in step potentials}
\author{Matthias Uhl, Volker Weissmann and Udo Seifert}
\address{II. Institut f\"ur Theoretische Physik, Universit\"at Stuttgart, 70550
Stuttgart, Germany}
\date{\today}

\begin{abstract}
	Although driven Brownian particles are ubiquitous in stochastic dynamics and
	often serve as paradigmatic model systems for many aspects of stochastic
	thermodynamics, fully analytically solvable models are few and far between.
	In this paper, we introduce an iterative calculation scheme, similar to the
	method of images in electrostatics, that enables one to obtain the
	propagator if the potential consists of a finite number of steps. For the
	special case of a single potential step, this method converges after one
	iteration, thus providing an expression for the propagator in closed form. In
	all other cases, the iteration results in an approximation that holds for
	times smaller than some characteristic timescale that depends on the number
	of iterations performed. This method can also be applied to a related class
	of systems like Brownian ratchets, which do not formally contain step
	potentials in their definition, but impose the same kind of boundary
	conditions that are caused by potential steps.
\end{abstract}

\section{Introduction}%
\label{sec:introduction}
Diffusion driven by an external force is a paradigmatic example of a
stochastic process~\cite{gardiner_stochastic_2009} and has become a
testing ground for stochastic thermodynamics in
theoretical~\cite{seifert_stochastic_2012} as well as in experimental
studies~\cite{ciliberto_experiments_2017}.
Analytically solvable models, however, have remained elusive with only few
exceptions like flat potential landscapes with various types of boundary
conditions and harmonic potentials. This state of affairs stands in stark
contrast to the situation of
other more established fields of physics that deal with conceptually similar
equations.  Every physics undergraduate is taught how to solve the Schrödinger
equation for square potentials in order to gain some intuition for systems that
behave similarly. The Fokker-Planck equation~\cite{risken_fokker-planck_1996}
does, from a mathematical point of view, not differ too much from the one
dimensional Schrödinger equation and yet there is no discussion of square
potentials or the like to be found that goes beyond the identification of the
stationary distribution (provided it exists due to appropriate boundary
conditions). In the present work, we will argue that it is indeed
worthwhile and feasible to seek for the time-dependent solution of the
Fokker-Planck equation with localized initial distribution, \ie{}, to determine the propagator
$p(x,t|\xinit,t=0)$ in one dimensional systems with forces arising from step
potentials. Not only should this close this gap in the literature, it might
even prove useful in a more directly applicable sense, since potential steps or
correspondingly sharply localized bursts of force could be encountered in
systems that exhibit strong entropic
barriers~\cite{burada_entropic_2008,kim_giant_2017}. Moreover,  we will see in
section~\ref{sec:example_application_brownian_ratchet} that even if there is
\emph{a priori} no delta-like force present in a system, certain boundary
conditions can be interpreted as if this was the case.

We first fix some notation and state the main objective of this work.
We are interested in solving the Fokker-Planck equation
\begin{equation}
	\partial_{t} p(x,t) = -\partial_{x} \left(\mu F(x) - D
	\partial_{x}   \right) p(x,t)
	\label{eq:fpEquation}
\end{equation}
for the probability distribution $p(x ,t)$ of a particle with mobility $\mu$
and diffusion coefficient $D=\mu \kb T$ with forces of the form
\begin{equation}
	F(x) = \f -\sum_{i=1}^{N}  \Delta V_{i} \delta(x-x_{i})
\end{equation}
arising from a driving force $\f$ and a potential with $N$ steps at arbitrary positions
$x_{i}$ and individual step heights of $\Delta V_{i}$. The initial distribution
is assumed to be localized at some point $\xinit$, \ie{}, $p(x,t=0) =
\delta(x-\xinit)$. In order to limit the number of parameters to a bare
minimum, we use reduced units. Quantities with the dimension of a length are
given as multiples of some characteristic length scale $\Delta x$ present in
the system, \eg{}, the distance between two specific potential steps. Time-like
quantities are expressed in multiples of the diffusion time $\tau \equiv \Delta
x^2/D$ and energies in multiples of the thermal energy $\kb T$. Using this
set of units, $\Delta x$, $D$, and $\mu$ are set to unity, leading to the
Fokker-Planck equation of the form $\partial_{t}p(x,t) = -\partial_{x}j(x,t)$
with the probability current defined as
\begin{equation}
	j(x, t) = (F(x) - \partial_{x})p(x, t)\,.
	\label{eq:probabilityCurrent}
\end{equation}

The remainder of this work is structured as follows.
In section~\ref{sec:boundary_conditions}, we derive the boundary conditions for
the time-dependent probability distributions on either side of a
potential step
from first principles. In section~\ref{sec:singlePotentialStep}, a
method for constructing the propagator following these boundary conditions is
introduced and subsequently applied to a system with a single potential step
and free boundary conditions. This results in
an analytical expression for the propagator in this system. We extend the scope
to systems with an arbitrary number of potential steps and periodic boundary
conditions in section~\ref{sec:multiplePotentialSteps}. As an example with
relevance in physics, in section~\ref{sec:example_application_brownian_ratchet} we discuss the propagator in the Brownian ratchet model
that is used to describe translocation of stochastically moving strands through
pores~\cite{mahendran_protein_2012,hepp_kinetics_2016,uhl_force-dependent_2018a}.

\section{Boundary Conditions}%
\label{sec:boundary_conditions}
For a single potential step, we expect, in analogy to the quantum mechanical counterpart, that the time-dependent
solution $p(x, t)$ is smooth everywhere except for the position of the
potential step $x_{1}$  and jumps at $x_{1}$, thus giving rise to the piecewise
defined solution
\begin{equation}
	p(x, t) = \left\{ \begin{array}{ll}
			p_{1}(x, t) & \text{for}\,x < x_{1}   \\
			p_{2}(x, t) & \text{for}\,x > x_{1}
	\end{array}  \right. \,.
	\label{eq:piecewiseProbability}
\end{equation}
The time-dependent
probability distributions within these half axes, which in view of later
generalizations will be called segments, are connected by
boundary conditions that are influenced by the step height. In this section, we derive
these boundary conditions from first principles.

The Fokker-Planck equation is a continuity equation for the probability
density. It describes how probability is redistributed over time and
the probability current $j(x, t)$ quantifies the amount of probability passing
through a point $x$ at time $t$. This interpretation should remain valid if
the potential is discontinuous at $x$, since in reality the discontinuity is
just a steep rise in potential. For this
reason, the probability current should not be discontinuous even if the
potential jumps at some point.

An immediate consequence of a continuous probability current across potential steps is
that the left- and right- sided limits of the current have
to agree when approaching the step position, \ie{},
\begin{equation}
	\lim_{x \nearrow x_{1}} j(x,t) = \lim_{x \searrow x_{1}} j(x,t)\,.
	\label{eq:current_limits}
\end{equation}
Within each interval, the current is given by~\eqref{eq:probabilityCurrent} with
only the constant force present. Therefore, eq.~\eqref{eq:current_limits}
translates into a boundary condition for $p(x, t)$ and its spatial derivative
\begin{equation}
	\f p_{1}(x_{1}, t) - \partial_{x}p_{1}(x, t) |_{x=x_{1} } = \f p_{2}(x_{1}, t) -
	\partial_{x}p_{2}(x, t)|_{x=x_{1}}\,.
	\label{eq:currentConditionProb}
\end{equation}

A further boundary condition that involves the characteristics of
the potential step can be derived by inverting the relation between current and
probability density. For, if the current $j(x, t)$  is known at some time $t$, the
probability density can be reconstructed through integration of
eq.~\eqref{eq:probabilityCurrent} via the relation
\begin{equation}
	p(x', t) e^{W(x, x')} = p(x, t) - \int_{x}^{x'}
	e^{W(x, x'')} j(x'', t) \,\mathrm{d}x'' \,,
	\label{eq:integratedProbabilityCurrent}
\end{equation}
where we denote the work needed for the transition from $x$ to $x'$ as
\begin{equation}
	W(x,x') \equiv -\int_{x}^{x'}F(x'') \, \mathrm {d}x'' \,.
	\label{eq:work}
\end{equation}
We now consider the limit of eq.~\eqref{eq:integratedProbabilityCurrent} for $x
\rightarrow x'$ with $x<x_{1}<x'$.
The integral on the right hand side is of order
$\mathcal{O}(x'-x)$ and the limit of the work from $x$ to $x'$ is just
the height $\Delta V$ of the potential step. Therefore, this limit yields
\begin{equation}
	p_{1}(x_{1}, t) = p_{2}(x_{1}, t)e^{\Delta V}   \,.
	\label{eq:probJumpCondition}
\end{equation}
Equation~\eqref{eq:probJumpCondition} implies that the potential step causes a
step in the opposite direction for the probability, which \textit{a posteriori}
is not too surprising
considering that a stationary Boltzmann-distribution in a periodic square potential
will show exactly this behavior. The crucial point is that this boundary
condition holds for a non-stationary dynamics as well.
Since the Fokker-Planck equation contains spatial derivatives up to second
order, boundary conditions for the probability density
(eq.~\eqref{eq:probJumpCondition}) and first derivative
(eq.~\eqref{eq:current_limits}) are sufficient to determine the evolution of
the probability density uniquely.

It is interesting to note that this treatment can be applied in the
reverse direction. If, for whatever reason, it is known that the probability
density jumps at a specific point by some factor and the probability current is
continuous, this behavior can be interpreted as if there was a potential step
present at this point.
This is similar to the treatment of resetting events that
can also be expressed within the formalism of Fokker-Planck equations by
augmenting the equation with terms representing the
resetting~\cite{manrubia_stochastic_1999, evans_diffusion_2011,
meylahn_large_2015, fuchs_stochastic_2016}.

\section{Single Potential Step}%
\label{sec:singlePotentialStep}

As we will see in later sections, solving the Fokker-Planck equation piecewise
with boundary conditions given in eq.~\eqref{eq:currentConditionProb}
and~\eqref{eq:probJumpCondition} can prove
quite challenging in the case of arbitrary step positions while a driving force
is applied. In order to introduce the key element of the solution technique, it
is instructive to start with the simplest
model, \ie{}, a single potential step $\Delta V$ without a driving
force and subject to free boundary conditions at infinity, \ie{}, pure
relaxation. Without loss of generality, we
assume that the step is located at $x_{1}=0$ and that the distribution is
initially localized at some point $\xinit<0$.

\subsection{Pure relaxation}%
\label{sub:noDrivingSingle}

For $f=0$, the Fokker-Planck equation
on the two halves of the real axis can be solved with the ansatz
\begin{align}
	\fl p_{1}(x,t) &=  \frac{1}{2 \sqrt{\pi t}}  e^{- \frac{(x-\xinit)^{2}  }{4t}  }
	+\frac{A}{2 \sqrt{\pi t}}  e^{- \frac{(x+\xinit)^{2}  }{4t}  } \equiv
	\frac{1}{2 \sqrt{\pi t}} \int_{-\infty}^{\infty} q_{1}(x') e^{- \frac{(x-x'
	)^{2} }{4t}}\,\mathrm{d}x'    \nonumber\\
	\fl p_{2}(x,t)	&=  \frac{B}{2 \sqrt{\pi t}}  e^{- \frac{(x-\xinit)^{2}  }{4t}
	} \equiv
	\frac{1}{2 \sqrt{\pi t}} \int_{-\infty}^{\infty} q_{2}(x') e^{- \frac{(x-x'
	)^{2} }{4t}}\,\mathrm{d}x' \,,
	\label{eq:singleStepAnsatz}
\end{align}
where $A$ and $B$ are constants yet to be determined.

\begin{figure}
	\centering
	\includegraphics[scale=1]{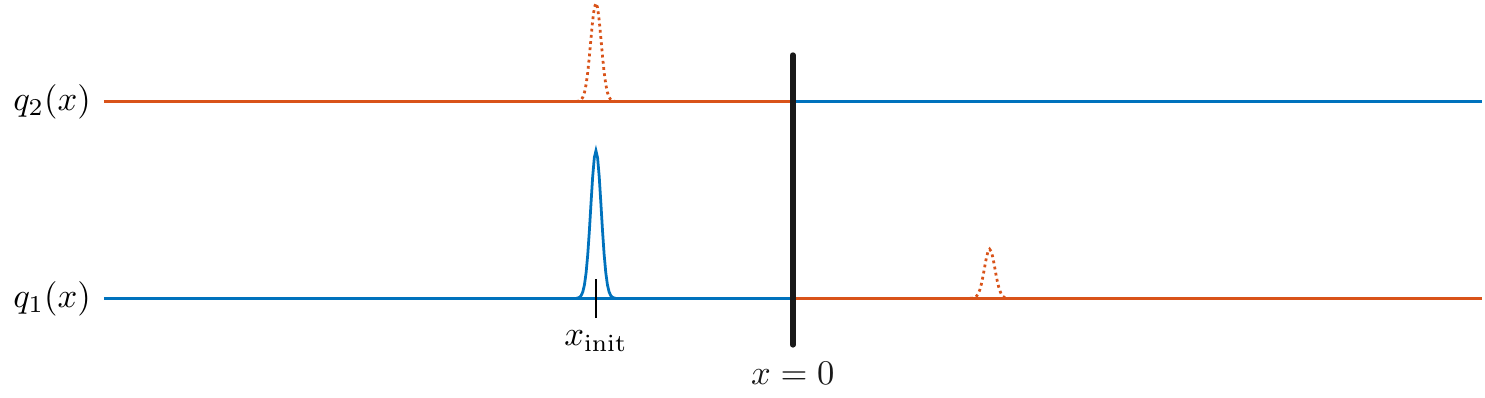}
	\caption{Sketch of the reflection scheme that leads to the
	analytical solution for the propagator in a system with free boundary
	conditions and a single potential step at $x_{1} =0$. The initial
distribution is reflected at the step position and multiplied by $\tanh(\Delta
V/2)$ to obtain the virtual initial distribution for the positive half axis. The
virtual initial distribution for the positive half axis is obtained by
multiplying the initial distribution with the factor $(1-\tanh(\Delta V/2))$.}%
	\label{fig:singleStepScheme}
\end{figure}

For $x<0$, the time evolution looks as if there was a mirror
image of the initial distribution present on the opposing side of the step. On
the positive half of the real axis, the probability distribution evolves like
coming from a localized initial condition, albeit with an modified amplitude.
These virtual initial distributions $q_{1}$ and $q_{2}$  are depicted in
figure~\ref{fig:singleStepScheme}.

An ansatz of this form is suitable, because by placing virtual initial
probability at the mirror position of the real initial probability, the ratio
of left and right sided probability and its derivative becomes
time-independent. Specifically,
\begin{equation}
	\frac{p_{1}(0,t) }{p_{2}(0,t) }  = \frac{1+A}{B} \quad \text{and} \quad
	\frac{\partial_{x}p_{1}(x,t)|_{x=0}}{\partial_{x}p_{2}(x,t)|_{x=0}}  =
	\frac{1-A}{B}
	\label{eq:singleStepConditions}
\end{equation}
for all times.
For $f=0$, the boundary conditions~\eqref{eq:currentConditionProb}
and~\eqref{eq:probJumpCondition} are of the same
type as these equations and satisfying them is now a simple matter of solving
for the constants $A$ and $B$ with the solution
\begin{equation}
	A = h \equiv \tanh(\Delta V/2) \quad \text{and} \quad B = 1-h\,.
	\label{eq:singleStepCoefficients}
\end{equation}

\subsection{Driven Brownian particle}%
\label{sub:driven_brownian_particle}

With
driving, \ie{}, $f\neq0$, the condition of current
continuity~\eqref{eq:current_limits} cannot be
fulfilled by the simple ansatz used in section~\ref{sub:noDrivingSingle}. In this section we
will demonstrate, however, that it is still possible to solve the Fokker-Planck
equation and the boundary conditions by introducing virtual initial probability
distributions that we denote by $q_{i}(x)$  for both intervals, however of a
more involved form than a simple mirror image of the real initial condition
with modified amplitude.

In analogy to eq.~\eqref{eq:singleStepAnsatz}, our
ansatz for the propagator reads
\begin{equation}
	p_{i} (x, t) = \frac{1}{2 \sqrt{\pi t}}   \int_{-\infty}^{\infty}
	q_{i}(x') e^{- \frac{(x-x' - \f t)^{2} }{4t}}\,\mathrm{d}x'\,,
	\label{eq:virtualProbabilityAnsatz}
\end{equation}
which corresponds to the time evolution of a virtual initial probability
distribution $q_{i}(x)$  for
each segment $i=1, 2$. Note that $q_{i}(x)$ is defined for all real numbers and is not
restricted to the interval for which $p_{i}(x,t)$ is defined. Furthermore, the
virtual initial distribution does not adhere to the rules normal probability
distributions must follow. It does not need to be normalized and can become
negative.
For the total probability distribution these properties usually are enforced
by the boundary conditions connecting the segments to each other and the fact
that the distribution is initially normalized and non-negative.

We impose that the $q_{i}(x)$ lead to the correct initial
distributions $p_{i}(x,t=0)$, which means that $q_{1}(x) = p_{1}(x,t=0)$ for
all $x<0$ and $q_{2}(x) = p_{2}(x, t=0)$ for $x>0$. The segments of $q_{i}(x)$
outside of the respective intervals, here, the other half axis, have to be determined such that the boundary
conditions between neighboring segments are respected by the resulting
probability distribution.

Using the ansatz
\begin{equation}
	q_{i}(x) \equiv g_{i}(x)e^{\f x/2}
	\label{eq:scaling}
\end{equation}
for the virtual
initial distribution, the time evolution can be rewritten in the form
\begin{equation}
	\fl p_{i}(x, t) = \frac{e^{-\f^{2}t/4 } }{2 \sqrt{\pi t}}
	\int_{-\infty}^{\infty} g_{i}(x')e^{\f x/2} e^{- \frac{(x-x')^2}{4t} }
	\,\mathrm{d}x' = %
	c(t)
	\int_{-\infty}^{\infty} g_{i}(z+x)e^{\f x/2}  e^{- \frac{z^2}{4t}  }
	\,\mathrm{d}z
	\label{eq:virtualProbabilityTransform}
\end{equation}
with the time-dependent prefactor $c(t)\equiv{e^{-\f^{2}t/4}}/{2\sqrt{\pi t}}$
and $z\equiv x-x'$.
By splitting the integration interval at zero and substituting
$u\equiv z^2$, eq.~\eqref{eq:virtualProbabilityTransform} can be rewritten in
terms of a Laplace transformation
\begin{align}
	p_{i}(x,t) &= c(t) e^{\f x/2}  \int_{0}^{\infty} \frac{1}{2 \sqrt{u}} \left(
	g_{i}(x+\sqrt{u}) +g_{i}(x-\sqrt{u}) \right) e^{- \frac{u}{4t}}
	\mathrm{d}u \nonumber\\
			   &= c(t) e^{\f x/2} \mathcal{L}_{u}  \left[ \bar{g}_{i}(x,
				   \sqrt{u}) /\sqrt{u}
		   \right] \left( \frac{1}{4t}   \right)\,.
	\label{eq:virtualProbabilityLaplace}
\end{align}%
We use the notation
\begin{equation}
	\mathcal{L}_{u}[f(u)](z) \equiv \int_{0}^{\infty} f(u) e^{-zu} \mathrm{d}u    
\end{equation}
and
we introduce the symmetric part of $g_{i}(z)$ with respect to $x$ as
\begin{equation}
	\bar{g}_{i}(x,z) \equiv (g_{i}(x+z) + g_{i}(x-z))/2\,.
	\label{eq:symmetrization}
\end{equation}
Along similar lines, the spatial derivative of the distribution can be
expressed as
\begin{equation}
	  \partial_{x} p_{i}(x,t)
	  = c(t)e^{\f x/2} \mathcal{L}_{u} \left[ \left(
			  \partial_{x} \bar{g}_{i}(x,\sqrt{u})   + \frac{\f}{2}
			  \bar{g}(x,\sqrt{u})
	  \right)/\sqrt{u}  \right] \left( \frac{1}{4t}   \right)
	  \label{eq:derivative}
\end{equation}
Due to the uniqueness of the Laplace transformation, the boundary
conditions~\eqref{eq:probJumpCondition} and~\eqref{eq:current_limits} that must
hold for all times transform into conditions on $\bar{g}_{i}(x,z)$ that must
hold for all $z$, \ie{},
\begin{equation}
	\bar{g}_{1}(x_{1},z) =
	e^{\Delta V} \bar{g}_{2}(x_{1},z)
	  \label{eq:jumpCondition}
\end{equation}
and
\begin{equation}
	\frac{\f}{2}\bar{g}_{1} (x_{1}, z)
	- \partial_{x}  \bar{g}_{1}(x, z)|_{x=x_{1} }
	= \frac{\f}{2}\bar{g}_{2} (x_{1}, z)
	-\partial_{x}  \bar{g}_{2}(x, z)|_{x=x_{1} }
	\,.
	\label{eq:currentCondition}
\end{equation}

Initially, the functions $g_{i}(x)$ are known only on the respective half axis
where they have to correspond to $p(x ,t)$, so
for equations~\eqref{eq:jumpCondition} and~\eqref{eq:currentCondition} to be
useful, they have to be rewritten such that they allow the calculation of
$g_{i}(x)$ on the other half axis. It
turns out that this goal can be achieved by assuming that the functions $g_{1}
(x_{1} - z)$ and $g_{2}(x_{1} +z)$ are known for positive $z$ and solving for the yet unknown
$g_{1}(x_{1}+z)$ and $g_{2}(x_{1}+z)$. This scheme leads to the uncoupled
inhomogeneous differential equations
\begin{equation}
	g'_{1}(x_{1} +z) - \frac{\f h}{2} g_{1}(x_{1} +z) = \psi(z)
	\label{eq:odeRight}
\end{equation}
and
\begin{equation}
	g'_{2}(x_{1}-z) + \frac{\f h}{2}  g_{2}(x_{1} -z) = \phi(z)
	\label{eq:odeLeft}
\end{equation}
with the inhomogeneous terms
\begin{equation}
	\psi(z) \equiv \frac{h \f }{2}  g_{1}(x_{1} -z)  - h
	g'_{1}(x_{1}-z)   + (1+h) g'_{2}(x_{1} +z)
	\label{eq:inhomoRight}
\end{equation}
and
\begin{equation}
	\phi(z) \equiv -\frac{h \f}{2}  g_{2}(x_{1} +z)
	+h g'_{2}(x_{1}+z) + (1-h) g'_{1}(x_{1} -z)\,,
	\label{eq:inhomoLeft}
\end{equation}
respectively. These can be readily solved by integration provided the unknown
functions are known for one reference
point $\xseam$. Thus we have found the desired linear operations that replace
the act of reflecting the initial
probability density in the case of non-vanishing driving force, namely
\begin{equation}
	\fl g_{1}(x) = \opSp[](\xseam) \left( \begin{array}{c}
			g_{1}(x) \\
			g_{2}(x)
	\end{array} \right) \equiv  g_{1}(\xseam) e^{h \f (x-\xseam)/2 } + \int_{\xseam}^{x}
	e^{- h \f(x' -x) /2} \, \psi(x'-x_{1})  \, \mathrm{d}x'
	\label{eq:solutionRight}
\end{equation}
and
\begin{equation}
	\fl g_{2}(x) = \opSm[](\xseam)   \left( \begin{array}{c}
			g_{1}(x) \\
			g_{2}(x)
		\end{array} \right) \equiv  g_{2}(\xseam) e^{-h \f (x-\xseam)/2 } +\int_{\xseam }^{x} e^{+h \f
	(x' -x) /2 } \, \phi(x_{1}-x')     \, \mathrm{d}x'\,.
	\label{eq:solutionLeft}
\end{equation}
Of course, the rescaled virtual initial distributions appear on both sides of
eqs.~\eqref{eq:solutionRight} and~\eqref{eq:solutionLeft} and so at first glance
they only seem to express a self consistency relation between the two
functions. The operators $\hat{S}^{\pm}$ are, however, constructed in such a
way that they can be evaluated if only parts of the functions are known.
If the driving force vanishes, the integrations
eqs.~\eqref{eq:solutionRight} and~\eqref{eq:solutionLeft} become a simple
reflection and thus
reproduce the result for this case we discussed above, as it is to be expected.

For a single potential step with driving force $f\neq0$ the situation is
as follows: Since the initial distribution is known, we have $q_{1}(x)
= \delta(x-x_{\mathrm{init}})$ for $x\leq0$, and  $q_{2}(x) =0$ for
$x \geq0$. These known function parts correspond to the blue intervals shown
in figure~\ref{fig:singleStepScheme}. The unknown parts, \ie{}, $q_{1}(x)$ for $x>0$
and $q_{2}(x)$ for $x<0$ can be obtained using the integration
scheme~\eqref{eq:solutionRight} and~\eqref{eq:solutionLeft}, respectively,
with the reference value $\xseam=0$ and $q_{1}(0) =
q_{2}(0) = 0$ is already known from the initial condition.

Performing the integrals and reversing eq.~\eqref{eq:scaling} yields
\begin{equation}
	\fl \pseudoP_{1}(x) = \delta(x-\xinit) + h e^{-\f \xinit }
	\delta(x+\xinit) + \theta(x+\xinit)
	\frac{\f}{2} e^{\f \left[ h (x+\xinit) +x -\xinit \right] /2 }  \left( h^{2} + h \right)
	\label{eq:singleStepInitLeft}
\end{equation}
and
\begin{equation}
	\fl \pseudoP_{2}(x) = (1-h)\delta(x-\xinit) +
	\theta(\xinit-x)\f e^{\f \left[ h (\xinit-x) +x-\xinit \right]/2} \left( h^{2}-h	  \right)/2
	\label{eq:singleStepInitRight}
\end{equation}
The time evolution of these virtual initial distributions according to
eq.~\eqref{eq:virtualProbabilityAnsatz} yields the exact
propagator for driven diffusion over a single potential step, which constitutes
one of our main results. We obtain
\begin{align}
	\fl p(x<0,t|\xinit,0) = \frac{1}{2 \sqrt{\pi t } } e^{-
		{(x-\xinit- \f t)^2 }/{4t} }+ \frac{ h e^{-\f \xinit } }{2
		\sqrt{\pi t}} e^{ -{(x + \xinit -\f t)^{2} }/{4t} } \nonumber\\
		\fl \!\! \qquad\quad + \frac{\f  e^{\f \xinit h } }{4} \left( h^{2} + h  \right)
		\left( \erf \left( \frac{\f t h + x+ \xinit }{2 \sqrt{t}} \right) +1
		\right)e^{ {\f } (x-\xinit) (1+h)/{2} - {\f^{2}t } (1-h^{2})/{4} }
	\label{eq:singleStepPropLeft}
\end{align}
and
\begin{align}
	\fl p(x>0,t|\xinit,0) = \frac{1-h}{2 \sqrt{\pi t} } e^{-
	{(x-\xinit-\f t )^2}/{4t} } \nonumber\\
	  \fl \qquad \quad+ \frac{\f}{4} (h-h^{2} ) \left( \erf \left( \frac{\f t h -
		 x+\xinit }{2 \sqrt{t} } \right) +1 \right) e^{ {\f}
		 (x-\xinit)(1-h)/{2 } -
	{\f^{2}t } (1-h^2)/{4} }\,.
	\label{eq:singleStepPropRight}
\end{align}

\begin{figure}[tpb]
	\centering
	\includegraphics[scale=1]{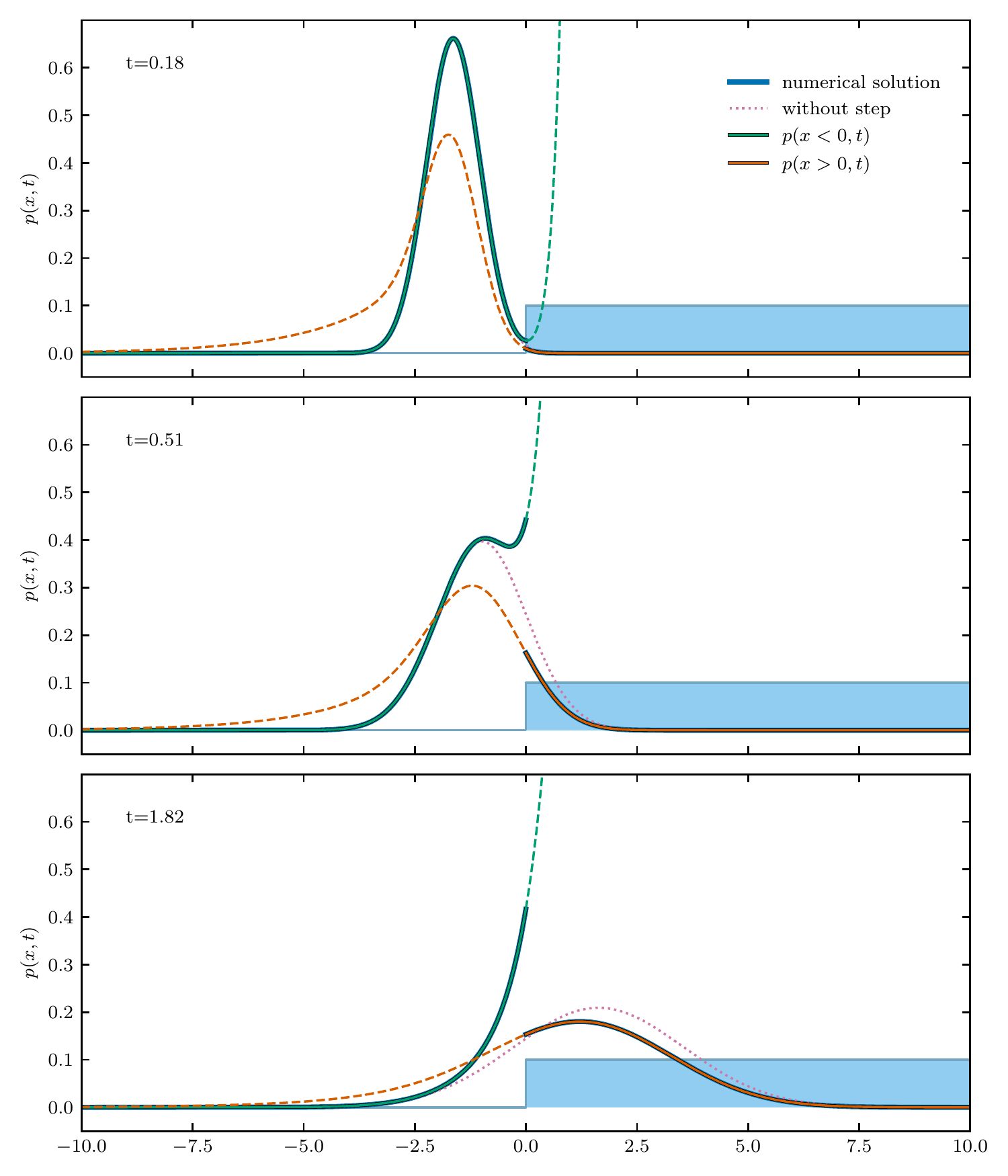}
	\caption{Propagator in a potential landscape with a single step of height
	$\Delta V=1$ located at $x_{1}=0$ and driving force $f=2$ to the right for different times. Shown are the time
evolutions of the virtual initial probabilities for the left and right half
axis (green and orange) in comparison to results obtained by numerical
integration of the Fokker-Planck equation and the propagator that would hold if
the potential step were not present (dotted). The virtual distributions are
drawn as solid lines in regions where they correspond to the propagator and as
dashed lines everywhere else. The analytical solutions shows excellent
agreement with the numerical one. Comparing with the bare diffusion/drift case, the step
leads to a delay of the peak position of the distribution and to broadening.}%
	\label{fig:singleStep}
\end{figure}

The time evolution for parameters $\f=2$ and $\Delta V = 1$ is shown in
figure~\ref{fig:singleStep} together with the results of a numerical solution
and the solution one expects without a potential step, \ie{}, for $\Delta V=0$.
For small times, when no significant portions of the distribution have reached
the step yet, the distribution looks Gaussian, as is to be expected.
When the broadening distribution reaches the step, parts of it are reflected,
leading to a diminished probability current when compared to the free
solution. This leads to the formation of a inflection point in the
probability distribution, as it is shown in the middle panel of
figure~\ref{fig:singleStep}. In the course of time most of the distribution crosses the
barrier and the distribution almost regains its Gaussian shape on the positive
half axis, although the distribution is slightly asymmetric with a longer tail
on the left than on the right. On the left half axis, the distribution transforms
into an almost exponential one, which corresponds to
the stationary distribution if the potential step was infinitely high (\cf{}
bottom panel). For even longer times, not shown in
figure~\ref{fig:singleStep}, all probability eventually crosses the potential
barrier and the time evolution of eq.~\eqref{eq:singleStepPropRight} (orange
curves) covers the full propagator in the long time limit.

\subsection{Reflecting and absorbing boundary conditions}%
\label{sub:reflecting_and_absorbing_boundary_conditions}

If the height of the potential step diverges, transitions across the step
position $x_{1}$ are possible only in one direction depending on the sign of
$\Delta V$. We can exploit this correspondence to determine the propagator of a
driven Brownian particle confined to the negative half axis, \ie{}, with
reflecting boundary conditions at $x_{1}=0$. This boundary condition
corresponds to $\Delta V \rightarrow \infty$ leading to $h=1$ in
eq.~\eqref{eq:singleStepPropLeft} that now simplifies to
\begin{align}
	p_{\text{ref.}}(x,t|\xinit, 0) &= \frac{1}{2 \sqrt{\pi t } } e^{-
		{(x-\xinit- \f t)^2 }/{4t} }+ \frac{  e^{-\f \xinit } }{2
		\sqrt{\pi t}} e^{ -{(x + \xinit -\f t)^{2} }/{4t} } \nonumber\\
								   &\qquad\quad + \frac{\f}{2}
		\left( \erf \left( \frac{\f t + x+ \xinit }{2 \sqrt{t}} \right) +1
		\right)e^{ {\f } x}
\end{align}
for $x<0$.
In the long-time limit the first two terms vanish and the third converges to
$p_{\text{eq}}(x)=\f e^{\f x}$ for positive $\f$, \ie{}, to the Boltzmann-distribution, as it is to
be expected.

If we have $\Delta V \rightarrow -\infty$ instead, no transitions are allowed
from the positive to the negative half axis while transitions in the opposing
direction are allowed. If the initial distribution is solely located on the
negative half axis, these boundary conditions have the same effect on the
negative half axis as absorbing boundary conditions. When inserting
$h=-1$ into~\eqref{eq:singleStepPropLeft} one obtains
\begin{equation}
	\fl p_{\text{abs.}} (x ,t|\xinit,0) = \frac{1}{2 \sqrt{\pi t } } e^{-
		{(x-\xinit- \f t)^2 }/{4t} }- \frac{e^{-\f \xinit } }{2
		\sqrt{\pi t}} e^{ -{(x + \xinit -\f t)^{2} }/{4t} } \\
\end{equation}
for $x<0$.

\section{Multiple potential steps}%
\label{sec:multiplePotentialSteps}
In contrast to a single step with free boundary conditions, an arbitrary number
of steps with possibly periodic boundary conditions, \ie{}, effectively an
infinite number of repeating steps, poses additional challenges to the
reflection scheme we introduced in section~\ref{sec:singlePotentialStep}. In
this section, we will present a more generalized strategy to calculate an
approximation to the propagator for these systems.

\subsection{Method of reflected probability}%
\label{sub:definitions}

When dealing with multiple potential steps, it is reasonable to split the
propagator at the positions of steps in the potential landscape. By using an
ansatz of the form
\begin{equation}
	p(x,t)  = \left\lbrace \begin{array}{ll}
			p_{1}(x,t) & \text{for}\,x \in (x_{0},x_{1})  \\
			p_{2}(x,t) & \text{for}\,x \in (x_{1},x_{2})  \\
			\cdots\\
			p_{i}(x,t) & \text{for}\,x \in (x_{i-1},x_{i})  \\
			\cdots\\
			p_{N+1}(x,t) & \text{for}\, x\in(x_{N},x_{N+1})
	\end{array} \right.\,,
	\label{eq:piecewiseProbabilityMult}
\end{equation}
where $x_{0}$ or $x_{N+1}$ might formally be set to positive or negative
infinity to indicate free boundary conditions.
The boundary conditions connecting one such segment $p_{i}(x ,t)$ to its
neighbors can be derived in the same way as it was performed for the single
step in section~\ref{sec:boundary_conditions}. The only difference is that the
resulting set of conditions take into account the properties of the individual
potential step, \ie{}, the position $x_{i}$ and the step height $\Delta V_{i}$.

As it was the case for the single step in
section~\ref{sub:reflecting_and_absorbing_boundary_conditions} one could
incorporate reflecting or absorbing boundary conditions, by
performing the limit $\Delta V_{i} \rightarrow \pm \infty$ at the first or last
step position. In any other position, a diverging step corresponds to a
boundary that allows transitions only in one direction.

Periodic boundary conditions that may be enforced in addition to
the conditions resulting from potential steps are mathematically equivalent to
the latter (eqs.~\eqref{eq:currentConditionProb}
and~\eqref{eq:probJumpCondition}) when setting the step height to zero and
using different values at which the respective functions are evaluated on the
left- and right-hand side of these equations. This relation between the two
types of
boundary conditions allows us to treat them both in a unified way by
introducing $p_{l(i)}$ and $p_{r(i)}$ for the segments of the
propagator to the left or right of the $i$-th potential step and the notation
$x_{l(i)}$ and $x_{r(i)}$ for the positions of these two functions that are
connected by the condition.
The unified boundary conditions read
\begin{equation}
	p_{l(i)}(x_{l(i)},t) = p_{r(i)}(x_{r(i)},t)e^{\Delta V_{i}}   \,.
	\label{eq:probJumpConditionMult}
\end{equation}
and
\begin{equation}
	\f p_{l(i)}(x_{l(i)},t) - \partial_{x}p_{l(i)}(x,t) |_{x=x_{l(i)} } = \f
	p_{r(i)}(x_{r(i)},t) -
	\partial_{x}p_{r(i)}(x,t)|_{x=x_{r(i)}}\,.
	\label{eq:currentConditionProbMult}
\end{equation}

If periodic boundary conditions are enforced, we may for simplicity assume,
without loss of generality, that $x$ is confined in the interval $[0, L]$,
where $L$ is the period length and that one of the potential steps is situated
at the periodic boundary, thus combining the boundary conditions of one step
with the periodic boundary condition.

\begin{figure}[tpb]
	\centering
	\includegraphics[width=0.8\linewidth]{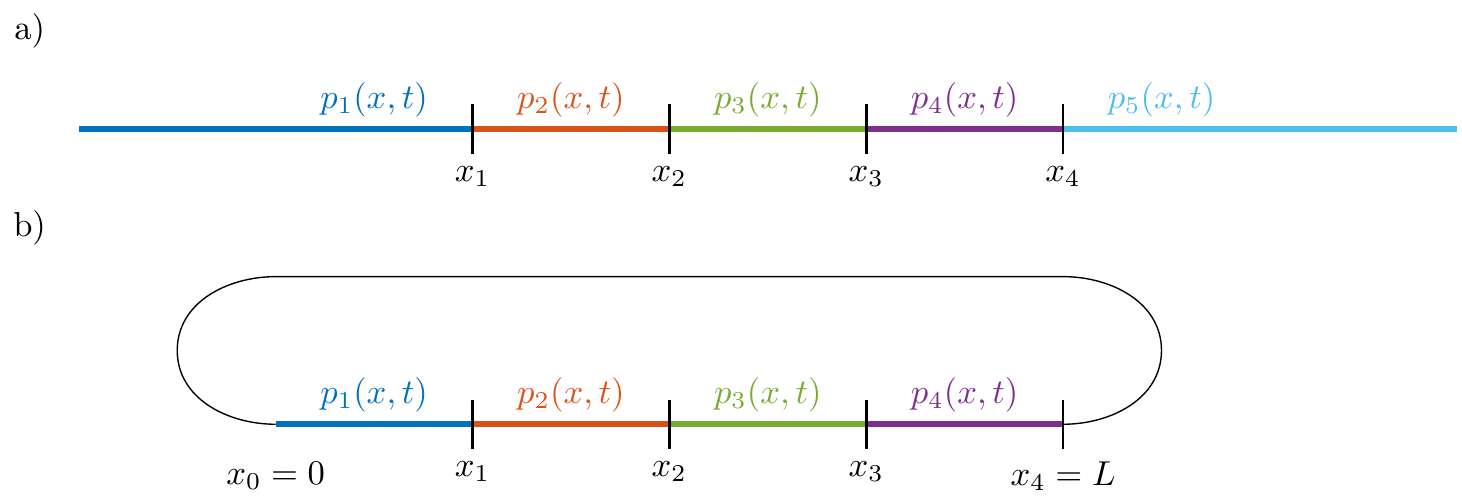}
	\caption{Illustration of step positions and enumeration of the segments
		$p_{i}(x,t)$ of the propagator in case of free (a) and periodic (b)
	boundary conditions. Vertical bars mark the positions of potential steps.
	With free boundary conditions $N$ steps in the potential split the domain
	in $N+1$ intervals, of which the first and last extend to positive and
	negative infinity, respectively (top panel). If periodic boundary
	conditions are imposed (bottom panel), we can, without loss of generality,
assume that one potential step is located at the boundary that connects the
first and the last segment.}%
	\label{fig:multi_step_periodic}
\end{figure}

In summary, the propagator is split into segments as shown in
figure~\ref{fig:multi_step_periodic}. For free boundaries (top panel), we use
an ansatz of the form of eq.~\eqref{eq:piecewiseProbabilityMult} with
$x_{0}=-\infty$ and $x_{N+1}=\infty$ with $N$ sets of boundary conditions of
the form of eq.~\eqref{eq:probJumpConditionMult} and
eq.~\eqref{eq:currentConditionProbMult}, where $p_{l(i)}=p_{i}$,
$p_{r(i)}=p_{i+1}$, and $x_{l(i)}=x_{r(i)}=x_{i}$.
Just as we have seen in
section~\ref{sub:reflecting_and_absorbing_boundary_conditions} it is also
possible to incorporate reflecting or absorbing boundary conditions by setting
$h_{i}=\pm 1$ on the outermost step positions.

For periodic boundary conditions (bottom panel), we formally have
$x_{0}= 0$ and $x_{N}=L$ in the ansatz~\eqref{eq:piecewiseProbabilityMult} that
is now missing the last segment. For all potential steps $i<N$ the boundary
conditions are the same as for the free boundary case. For the singular
transition $i=N$, we have $p_{l(N)}=p_{N}$, $p_{r(N)}=p_{1}$, $x_{l(i)}=x_{N}$,
and $x_{r(i)}=x_{1}$.

After introducing these definitions, we can repeat the same steps that led to
the solution of the single step case. In doing so, we introduce the virtual
initial distributions corresponding to each segment of the propagator as
\begin{equation}
	\fl p_{i}(x ,t) =  \frac{1}{2 \sqrt{\pi t}}   \int_{-\infty}^{\infty}
	q_{i}(x') e^{- \frac{(x-x' - \f t)^{2} }{4t}}\,\mathrm{d}x' = 
	c(t) \int_{-\infty}^{\infty} g_{i}(z+x) e^{f_{0}x/2 } e^{-
	\frac{z^{2} }{4t}} \,\mathrm{d}z\,.
	\label{eq:virtualProbabilityAnsatzMult}
\end{equation}
In analogy to the single step, we can compute the unknown parts of the rescaled
virtual initial distributions by integrating the differential equations
\begin{equation}
	g'_{l}(x_{l} +z) - \frac{\f h_{i}}{2} g_{l}(x_{l} +z) = \psi_{l,r}(z)
	\label{eq:odeRightMult}
\end{equation}
and
\begin{equation}
	g'_{r}(x_{r}-z) + \frac{\f h_{i} }{2}  g_{r}(x_{r} -z) = \phi_{l,r}(z)
	\label{eq:odeLeftMult}
\end{equation}
with $h_{i}\equiv \tanh(\Delta V_{i}/2)$. The inhomogeneities read
\begin{equation}
	\psi_{l,r}(z) \equiv \frac{h_{i} \f }{2}  g_{l}(x_{l} -z)  - h_{i}
	g'_{l}(x_{l}-z)   + (1+h_{i}) e^{\f (x_{r}-x_{l})/2} g'_{r}(x_{r} +z)
	\label{eq:inhomoRightMult}
\end{equation}
and
\begin{equation}
	\phi_{l,r}(z) \equiv -\frac{h_{i}\f }{2}  g_{r}(x_{r} +z)
	+h_{i}g'_{r}(x_{r}+z) + (1-h_{i}) e^{\f (x_{r}-x_{l})/2  } g'_{l}(x_{l}
	-z)\,,
	\label{eq:inhomoLeftMult}
\end{equation}
where we have suppressed the $i$-dependence of $r(i)$ and $l(i)$ for
notational simplicity.
The solutions are
\begin{equation}
	\fl g_{l}(x) = \opSp(\xseam) \left( \begin{array}{c}
			g_{l}(x) \\
			g_{r}(x)
		\end{array} \right) \equiv  g_{l}(\xseam) e^{h_{i} \f( x- \xseam)/2 } + \int_{\xseam }^{x} e^{- h_{i}
	\f( x' -x) /2} \, \psi_{l,r}(x'-x_{l})  \, \mathrm{d}x'
	\label{eq:solutionRightMult}
\end{equation}
and
\begin{equation}
	\fl g_{r}(x) = \opSm(\xseam)   \left( \begin{array}{c}
			g_{l}(x) \\
			g_{r}(x)
		\end{array} \right) \equiv  g_{r}(\xseam) e^{-h_{i} \f (x-\xseam)/2 } +\int_{\xseam }^{x} e^{+h_{i} \f
	(x' -x) /2 } \, \phi_{l,r}(x_{r}-x')     \, \mathrm{d}x'\,,
	\label{eq:solutionLeftMult}
\end{equation}
respectively.

The main difference to the case of a single potential step is that it is no
longer possible to calculate the unknown parts of the virtual initial
distributions by applying the reflection operators $\opSp$ and $\opSm$ once.
We rather have to apply them repeatedly, each time generating yet unknown parts
of the virtual initial distributions.  If, for instance, $g_{l}(x)$
is known in the interval $I_{l}$  and $g_{r}(x)$ is known in the interval
$I_{r}$, then the integral in eq.~\eqref{eq:solutionRightMult} can be evaluated in
the interval $\intSp (\intr,\intl ) \equiv  (2 x_{l} -I_{l}) \cap (I_{r} -x_{r}
+x_{l})$, provided one reference value $g_{l}(\xseam)$ at the reference
position $\xseam$ within this interval is
known.  Equation~\eqref{eq:solutionLeftMult} can be evaluated in the interval
$\intSm(\intl, \intr) \equiv(2x_{r} -I_{r}) \cap( I_{l} -x_{l} + x_{r})$.
In the following, we will demonstrate that the reflection operators
$S_{i}^{\pm} $ in conjunction with the initial distribution is sufficient to
obtain the propagator in an iterative procedure. This iterative approach is
conceptually similar to the problem of finding the electrostatic potential for a charged
particle in between two conducting parallel surfaces often discussed in
electrodynamics courses, where the method of mirror charges yields a solution
in form of an infinite sum~\cite{jackson_classical_1999}.
The concrete strategy that has to be employed, however, will depend on the
specific potential landscape that is to be studied.

\subsection{Two steps with free boundary conditions}%
\label{sub:two_jumps_with_free_boundary_condtions}
To illustrate how to calculate the virtual initial distributions in more
complex situations, we first consider the case of a potential landscape with two
steps with height $\Delta V_{1}$ and $\Delta V_{2}$ at the positions $x_{1}$
and $x_{2}$, respectively.

\begin{figure}
	\centering
	\includegraphics[scale=1]{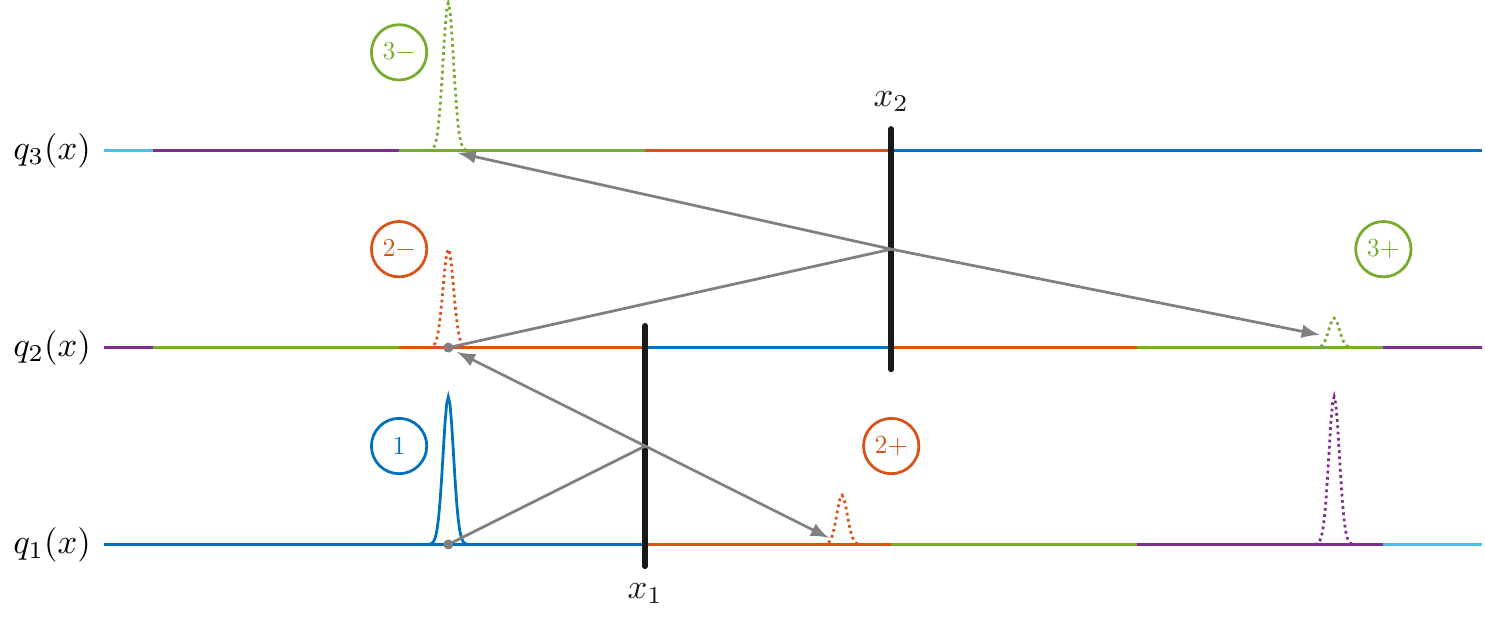}
	\caption{Sketch illustrating the reflection scheme for a potential with
	two steps placed at $x_{1}$ and $x_{2}$ and initial position placed to the
	left of $x_{1}$. Each horizontal line represents the virtual initial
	distribution responsible for a segment of the propagator. The vertical
	lines correspond to boundary conditions connecting adjacent segments.
	Initially, the virtual distributions are only known in the intervals where
	they correspond to the actual initial distribution. These intervals are
	marked in blue. By repeatedly applying the reflection operators, new parts of
	the virtual distributions are generated, which are colored in order orange,
	green, purple, and light blue.}%
	\label{fig:doubleStepSketch}
\end{figure}
Figure~\ref{fig:doubleStepSketch} depicts the three virtual initial distributions
$q_{i}$ relevant for the intervals $(-\infty,x_{1})$,$(x_{1},x_{2})$, and
$(x_{2},\infty)$ as horizontal lines and the potential steps between intervals as
vertical bars.
Initially, the virtual initial distributions are known only in those intervals
for which $q_{i}$ must correspond to the initial distribution. These intervals are marked in blue in
the figure. Applying the four operators $S_{i}^{\pm}$ with $i \in \{1,2\}  $
corresponding to the boundary conditions at the potential steps one can now
obtain parts of adjacent $q_{i}$ (marked in red) in analogy to the case of a single step. In
contrast to the single step, however, the initial interval in which $q_{2}$ is
known, has a finite length, and therefore the newly calculated parts are
limited by that length. To illustrate the procedure figure~\ref{fig:doubleStepSketch} depicts the case
without driving force for which the operators correspond to mere reflections. Here, we find that in this first step the initial
delta-peak (marked with 1) gets reflected by $S_{1}^{+}$ to a peak (marked as
$2+$) on the other side of the first step position $x_{1}$. Analogously,
$S_{1}^{-}$ maps the same peak with different height onto $q_{2}$ (marked as
$2-$).

When performing the scheme a second time, the peak $2-$ is reflected by the
operators $S_{2}^{\pm}$ with respect to the second step position $x_{2}$
resulting in new peaks on $q_{2}$ and $q_{3}$. The peaks are marked as $3+$ and
$3-$, respectively. In each following step, the peak that was generated on
$q_{2}$ in the step before, will give rise to a new peak on either $q_{1}$ or
$q_{3}$ and a new peak on $q_{2}$ (\cf{} purple peak in
fig.~\ref{fig:doubleStepSketch}, the new peak on $q_{2}$ lies outside of the depicted
interval).

Since each step of this scheme generates parts of the virtual initial
distribution only on intervals of finite length, it
would have to be repeated infinitely often to obtain the full functions
$q_{i}(x)$ for all real numbers. For practical purposes, it is
sufficient to stop after a finite number of iterations, which results in an
approximation valid for small times.
It is, however, necessary to specify how to treat the still unknown parts of the
virtual initial distributions when calculating the time evolution according to
eq.~\eqref{eq:virtualProbabilityAnsatzMult}. It turns out that, when driving forces
are present, the reflection operators may generate a virtual initial distribution that
grows exponentially with $x$ (\cf{} eq.~\eqref{eq:singleStepInitLeft}). So,
naively setting the virtual initial distribution to zero where it is unknown or
equivalently limiting the integration in
eq.~\eqref{eq:virtualProbabilityAnsatz} to the known parts of $q_{i}$ will
introduce an artificial step in the initial distribution that originates
solely from stopping the iteration after finitely many steps. To avoid this
step from occurring, we assume instead that the function that describes the
last known part of the function remains valid in the unknown intervals. While
this is not the correct initial distribution either, we observe that this
procedure, called \enquote{extension}, converges faster to the real propagator
than the naive procedure.

\begin{figure}
	\centering
	\includegraphics[scale=1]{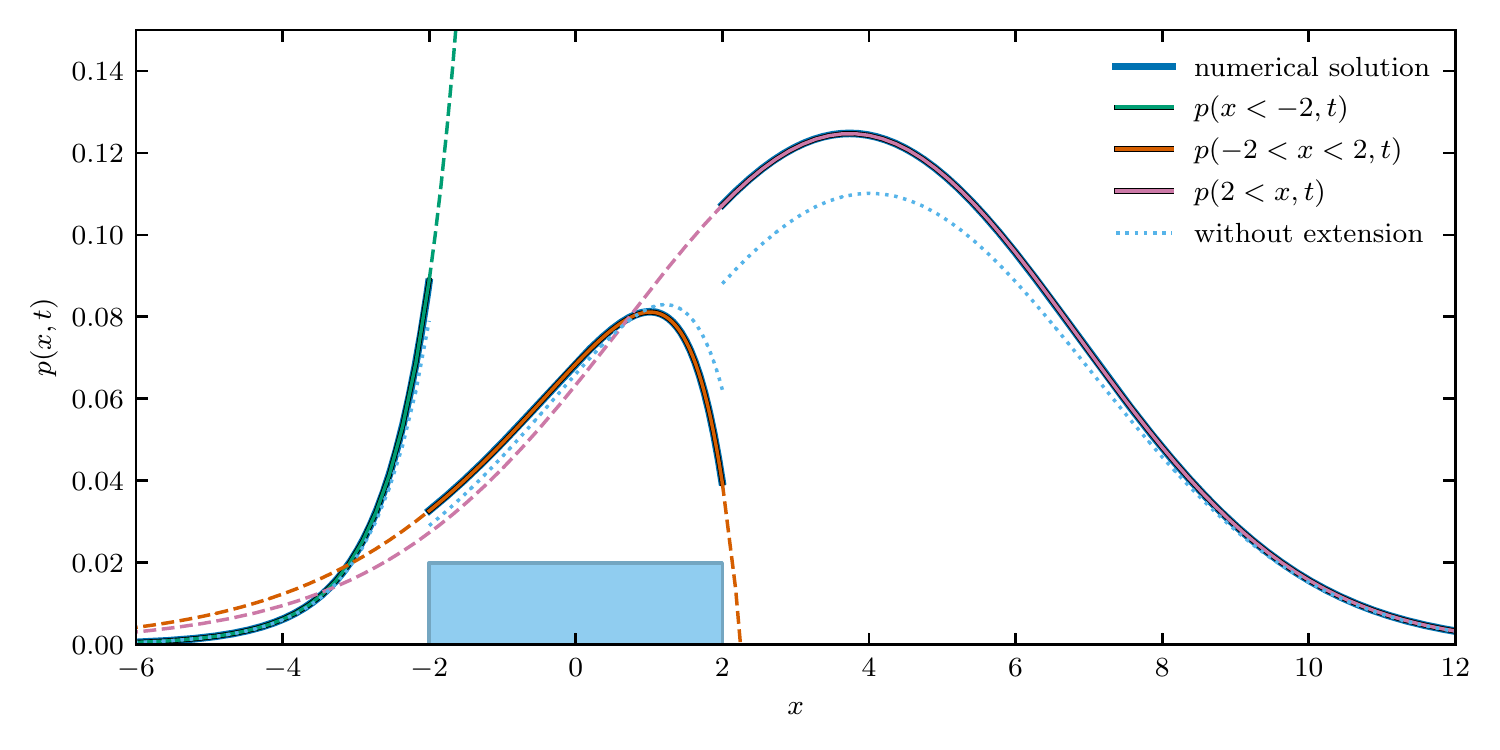}
	\caption{Propagator for an initially localized distribution at $\xinit =-5$
		driven by a force $\f=2$ in a potential landscape
		with step $\Delta V_{1}=1$ at $x_{1}=-2$ and step $\Delta V_{2}=-1$ (as
		indicated by the shaded rectangle) at
	$x=2$ at time $t=4.52$. The figure shows the piecewise defined approximation
	obtained by reflecting the initial distribution $N_{\mathrm{it}}=2$ times at the potential
	steps in comparison to a numerical solution of the Fokker-Planck
	equation showing excellent agreement despite the fact that the reflection
	scheme was performed only twice. If the left and
rightmost known intervals of the virtual initial distribution are not extended
to infinity the approximation becomes worse as is indicated by the
dotted lines. Without the extension, the iteration has to be repeated
$N_{\text{it}}=4$ times to achieve a match to the numerical solution within
error bars that correspond to the thickness of the curve (not shown).}%
	\label{fig:doubleStep}
\end{figure}

As an example, in Fig.~\ref{fig:doubleStep}, we show the propagator $p(x,t| x=-5,t=0)$
in a system with a rectangular potential barrier of height $\Delta V=1$ between $x=-2$ and $x=2$ for a
driving force of $\f=2$.

\subsection{Generalization to an arbitrary number of steps}%
\label{sub:arbitrary_number_of_steps}
The method discussed for the case of two potential
steps can straightforwardly be generalized to the case of an arbitrary number of
steps.
As before the reflection scheme will lead to an iterative solution.
We use the index $k$ to count iterations.
At the beginning of the iteration, the virtual distributions are known in the intervals
$I_{i}^{0} = [x_{i-1},x_{i}]$. The initial guess for the rescaled virtual
initial distribution is
\begin{equation}
	g_{i}^{0}(x) = \left\lbrace \begin{array}{lc}
		p(x,t=0) e^{-\f x/2}  & \text{for}\, x \in I_{i}^{0}  \\
		\text{undefined} & \text{otherwise}
	\end{array}  \right.\,.
\end{equation}
In each iteration step, one can generate new
parts for each $g_{i}$ by applying $\opSp[i](\xseam)$ and
$\opSm[i-1](\xseam)$ with a suitable choice for the seam position $\xseam$,
\ie{},
\begin{equation}
	g_{i}^{k+}(x) = \opSp(\xseam^{k+}) \left( \begin{array}{c}
			g_{i}^{k}(x) \\
			g_{i+1}^{k}(x)
	\end{array} \right)
\end{equation}
and
\begin{equation}
	g_{i}^{k-}(x) = \opSm[i-1](\xseam^{k-})  \left( \begin{array}{c}
			g_{i-1}^{k}(x) \\
			g_{i}^{k}(x)
	\end{array} \right)\,.
\end{equation}
For practical purposes it is advisable to choose $\xseam^{k+}$ as big as
possible within the known interval $I_{i}^{k}$ to avoid repeating operations
already performed in earlier iterations. For the same reason, $\xseam^{k-}$
should be chosen as small as possible.

The new functions $g_{i}^{k\pm} (x)$ generated in the $k$-th step can be
evaluated in the intervals $I_{i}^{k+} = \intSp[i](I_{i}^{k},I_{i+1}^{k})$ and
$I_{i}^{k-} = \intSm[i-1](I_{i-1}^{k},I_{i}^{k})$, respectively. These
intervals in all but the first step will have overlap with all previously
generated intervals but will also contain new parts. Since all generated
functions $g_{i}^{k}$ are constructed using the same operators, $g_{i}^{k\pm}$
will agree with the previous iterations in all parts that were already known.

These considerations lead to the updated interval
\begin{equation}
	I_{i}^{k+1} = I_{i}^{k} \cup I_{i}^{k+}   \cup I_{i}^{k-}.
\end{equation}
and the updated approximation of the rescaled virtual initial distribution
\begin{equation}
	g_{i}^{k+1}(x) = \left\lbrace \begin{array}{ll}
			g_{i}^{k}(x) & \text{for}\,x \in I_{i}^{k}     \\
			g_{i}^{k+}(x) & \text{for}\,x \in I_{i}^{k+} \setminus I_{i}^{k}   \\
			g_{i}^{k-}(x) & \text{for}\,x \in I_{i}^{k-}  \setminus I_{i}^{k}     \\
			\text{undefined} & \text{otherwise}
	\end{array} \right. \,.
\end{equation}
As stated above, excluding $I_{i}^{k}$ from the second and third case would not
strictly be necessary since the functions agree anyway.
Repeating this iteration scheme and computation of the time evolution according
to eq.~\eqref{eq:virtualProbabilityAnsatzMult} yields an approximation to the
desired propagator.

\subsection{An algorithm for a closed form of the propagator}%
\label{sub:feasibility_of_the_integration_procedure}
Requiring a large number of consecutive integrations of some functions
immediately poses the question, whether it is feasible to arrive at a result in
closed form.
While this most likely will not be possible for arbitrary initial distributions, we
will show in this section that  after an
arbitrary number of integration steps the virtual initial distribution can be expressed in closed form provided
that the initial probability is localized. Thus the propagator can be
expressed in closed form. For such an initial condition, the rescaled virtual
initial distribution can be expressed in the form
\begin{multline}
	 g_{i}(x)=\sum_{j} a_{i,j} \delta(x-\tilde{x}_{i,j} ) +
	\theta(x-\tilde{x}_{i,j}) \left( \sum_{k} \mathcal{P}_{i,j,k}^{+}
	(x)e^{b_{i,j,k}^{+} x } \right)\nonumber\\
	\quad + \theta(\tilde{x}_{i,j}-x) \left( \sum_{k}
\mathcal{P}_{i,j,k}^{-}(x) e^{b_{i,j,k}^{-}x}\right)\,,
\label{eq:virtualProbabilityUniversalForm}
\end{multline}%
where $\mathcal{P}^{\pm}_{i,j,k}(x)$ denotes some polynomial of $x$,
$\tilde{x}_{i,j}$ are positions of step- or delta-functions, $a_{i,j}$ are the
amplitudes of the delta-functions, and $b_{i,j,k}^{\pm}$ are real valued
coefficients in the exponential prefactors of the polynomials $\mathcal{P}^{\pm}_{i,j,k}(x)$.

Such an expression holds since the operations necessary to calculate the result
of the reflection operators defined in
eqs.~\eqref{eq:solutionRightMult} and~\eqref{eq:solutionLeftMult}, can, by means of
integration by parts, be broken down into a concatenation of the following
operations:
\begin{itemize}
	\item shifting in $x$: $g_{i}(x) \rightarrow g_{i}(x-c)$
	\item reflection of $g_{i}(x)$ at some point $c$: $g_{i}(x) \rightarrow
		g_{i}(2c-x)$
	\item multiplication with an exponential function: $g_{i}(x) \rightarrow
		 g_{i}(x) \, a e^{c x}$
	\item integration: $g_{i}(x) \rightarrow \int_{\xseam}^{x}
		g_{i}(x')\,\mathrm{d}x' $
	\item addition of multiple functions of the type of
		eq.~\eqref{eq:virtualProbabilityUniversalForm}.
\end{itemize}
Each of these five operations preserves the form of
eq.~\eqref{eq:virtualProbabilityUniversalForm} and can be executed in closed
form. Since the initial distribution is also of the same form, this guarantees
that an arbitrary number of reflection operations can be performed on the initial
distribution while still obtaining a result in closed form, \ie{}, of the form
of eq.~\eqref{eq:virtualProbabilityUniversalForm}. While the
calculation can become quite tedious, since it involves, among others,
recursive integration by parts and application of the binomial theorem, it is
possible to create an algorithm for performing these steps using standard computer algebra libraries.

Finally, the explicit calculation of the time evolution of the virtual initial
probabilities and, hence, the propagator, is also possible in closed form.
A reference implementation of such an algorithm is provided in the supplementary
material~\bibnote{\url{https://github.com/muhl/StepPotentialPropagator}}

\section{Brownian ratchet as an application}%
\label{sec:example_application_brownian_ratchet}

Driven diffusion in potentials with steps or, equivalently, driven diffusion with
boundary conditions of the form of eq.~\eqref{eq:probJumpConditionMult} and
eq.~\eqref{eq:currentConditionProbMult} are not only useful as introductory problems
to the topic of driven diffusion. They can be encountered, \emph{inter
alia}, in biophysical problems described as Brownian ratchets.
In this section, we will derive an approximation for the corresponding
propagator using the reflection scheme.

In brief, the Brownian ratchet model originally introduced by Oster and
Peskin~\cite{simon_what_1992,peskin_cellular_1993}
describes the stochastic dynamics of translocation of a polymer strand through a pore. It
is assumed that on one side of the pore molecules can bind to equidistant sites
on the strand that will block a retraction of the strand back through the
pore. The strand itself undergoes regular diffusion that may
be biased by a pulling force. The timescale of the binding and unbinding
processes of the bound molecules is assumed to be fast when compared to the
diffusion timescale, which means that one can assume that the binding sites are
always in chemical equilibrium with their respective environment. Formally,
this model
corresponds to the ensemble statistics of the distance $x$ of the barrier to the closest
binding site on one side of the pore with a Fokker-Planck of the form of
eq.~\eqref{eq:fpEquation}. This variable may range between $0$ when the
binding site is inside the pore to the distance of two binding sites $\Delta
x$.
\begin{figure}
	\centering
	\includegraphics[scale=1]{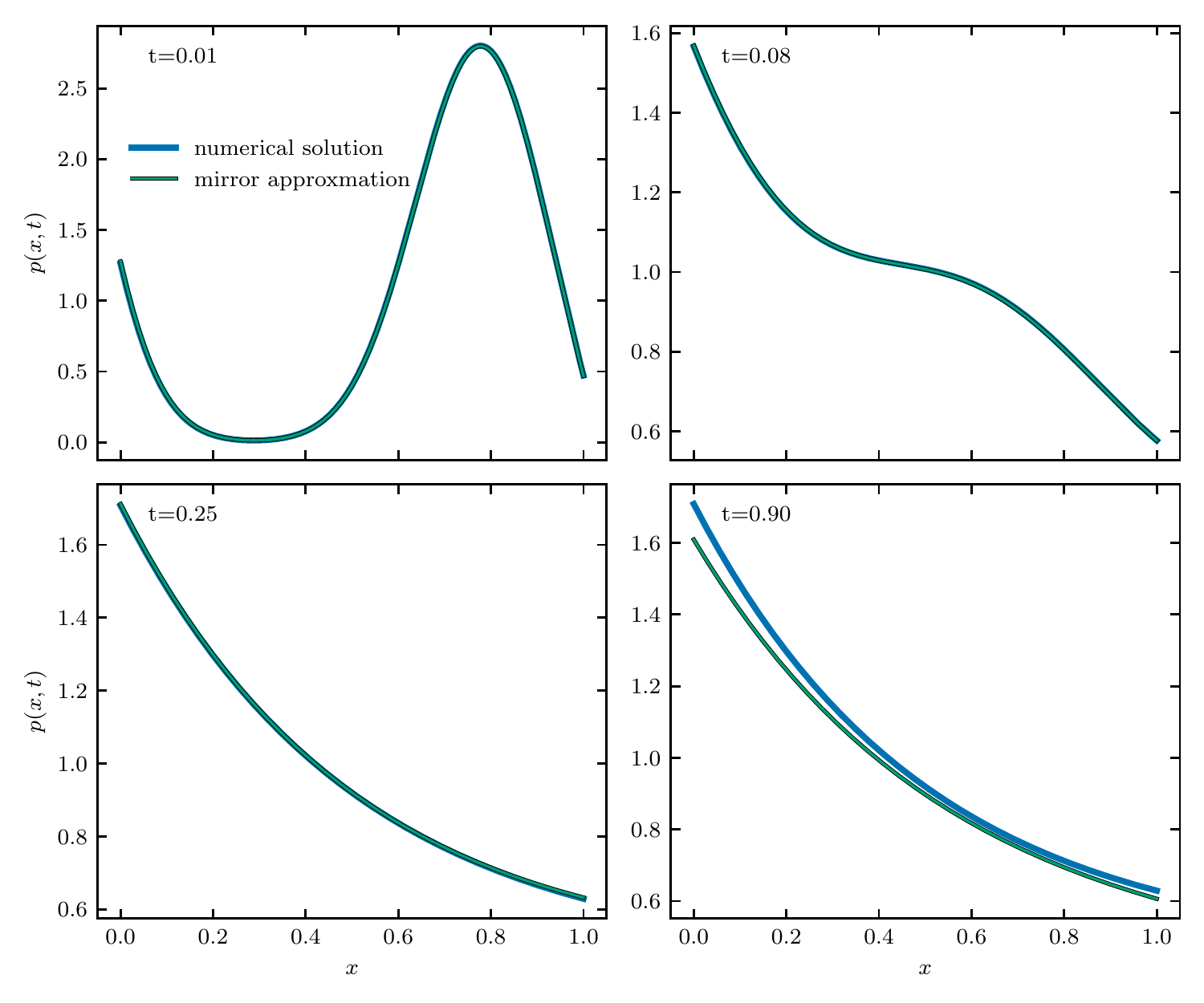}
	\caption{Time evolution of initial distribution localized at
	$\xinit=0.8$ in the Brownian ratchet system with $\Delta x = 1$. A pulling force of $\f =
	-2$ is applied. The probability to reject a transition across the periodic
	boundary is $1/e$ corresponding to a potential step with height $\Delta
	V=-1$. The plot compares the numerical integration of the Fokker-Planck
	equation to the approximation obtained using the reflection method with
	$N=2$ iterations. We find excellent agreement at early times (top row) and
	even until the stationary distribution is reached (bottom left). At even
	larger times the approximation starts to deviate from the numerical
	solution (bottom right) and is ultimately vanishing for $t \rightarrow
	\infty$.}%
	\label{fig:periodic_step}
\end{figure}
Transitions of $x=0$ to $x=\Delta x$ correspond to the transition of a
binding site back to the cis side and are therefore rejected with the stationary
probability $p_{\text{on}} = w_{\text{bind}}/(w_{\text{bind}} +
w_{\text{unbind}})$  of a binding site to be occupied. Transitions in the
opposite directions, \ie{}, the transition of an empty site from the trans to
the cis side are always allowed. These modified periodic boundary conditions
read
\begin{equation}
	\frac{p(0,t)}{p(\Delta x, t)}  = 1/p_{\text{on}} = e^{-\ln p_{\text{on}} }\,.
	\label{eq:brownianRatchetBoundary}
\end{equation}
To keep the probability distribution normalized at all times, the currents at
$x=0$ and $x=\Delta x$ have to match.
These boundary conditions are of the same form that arises from the presence of
potential steps. Comparison of eq.~\eqref{eq:brownianRatchetBoundary} and
eq.~\eqref{eq:probJumpConditionMult} shows that the height of the
potential step at the periodic boundary can be identified as $\Delta V = \ln
p_{\text{on}}$.

The propagator for this system can therefore be calculated with the reflection
scheme we discussed above. Because there is only one interval in this case,
the virtual initial distribution is obtained by repeatedly processing the
single virtual initial distribution with the operators $\opSp[]$ and $\opSm[]$
associated with the periodic boundary condition.

As a result the propagator for a initial distribution located at
$x=0.8$ is shown in figure~\ref{fig:periodic_step}. We see good agreement
between the numerical solution and an approximation with two
iterations well beyond the time at which the stationary exponential
distribution is reached.

\section{Conclusion}
In the present work, we have introduced a technique for constructing the propagator
in one-dimensional driven stochastic systems with discontinuous potential
landscapes. We have demonstrated that it is possible to satisfy the boundary
conditions that connect segments of constant potential by calculating the time
evolution of virtual initial distributions. This method is conceptually similar
to the well known method of images commonly used to solve problems involving
conducting surfaces in electrostatics.

While the technique will in all but the simplest example of a single potential
step involve an infinite recursion of mirror images, the virtual initial
distributions can nevertheless help to gain an intuitive picture of the
dynamics. If, as we have seen in section~\ref{sec:singlePotentialStep}, a
potential barrier blocks driven diffusion, the actual initial distribution
is opposed by a modified mirror image that represents the transitions blocked due to the barrier. Conversely, if a
negative step in potential draws probability across the step, the virtual
initial distributions may become negative.

It is possible to stop the reflection scheme after a
finite number of iterations and still gain a good approximation to the
propagator with significantly less computational effort than for a numerical
integration of the Fokker-Planck equation. Because the virtual initial
distributions are oblivious to periodic boundary conditions, the time
evolution predicted by them is unable to attain a non-vanishing stationary state.
However, since it is trivial to
compute the stationary state~\cite{gardiner_stochastic_2009}, a finite number of iterations is
sufficient to get an excellent approximation to the propagator for all times,
as we were able to demonstrate for the example of a Brownian ratchet.

Further research and optimization could be directed at the open question
whether the technique discussed here can be extended into an algorithm
suitable for more generic types of potentials. After all, every smooth
potential can be approximated by a sequence of potential steps. Preliminary
considerations show that the computational effort would scale not worse than
established integration schemes, \eg{}, finite difference approximations to
the Fokker-Planck equation~\cite{risken_fokker-planck_1996}, with the added benefit that discontinuities
in the potential can be handled without the risk of numerical instabilities.

\section*{References}%
\label{sec:references}

\bibliographystyle{iopart-num}
\bibliography{step_potential.bib}

\providecommand{\newblock}{}
\begin{thebibliography}{10}
\expandafter\ifx\csname url\endcsname\relax
  \def\url#1{{\tt #1}}\fi
\expandafter\ifx\csname urlprefix\endcsname\relax\def\urlprefix{URL }\fi
\providecommand{\eprint}[2][]{\url{#2}}

\bibitem{gardiner_stochastic_2009}
Gardiner C 2009 {\em Stochastic {{Methods}}: {{A Handbook}} for the {{Natural}}
  and {{Social Sciences}}\/} 4th ed Springer {{Series}} in {{Synergetics}}
  ({Berlin Heidelberg}: {Springer-Verlag}) ISBN 978-3-540-70712-7

\bibitem{seifert_stochastic_2012}
Seifert U 2012 {\em Rep. Prog. Phys.\/} {\bf 75} 126001

\bibitem{ciliberto_experiments_2017}
Ciliberto S 2017 {\em Phys. Rev. X\/} {\bf 7} 021051

\bibitem{risken_fokker-planck_1996}
Risken H 1996 {\em Fokker-{{Planck Equation}}\/} ({Springer Berlin Heidelberg})
  ISBN 978-3-540-61530-9 978-3-642-61544-3

\bibitem{burada_entropic_2008}
Burada P~S, Schmid G, Reguera D, Vainstein M~H, Rubi J~M and H{\"a}nggi P 2008
  {\em Phys. Rev. Lett.\/} {\bf 101} 130602

\bibitem{kim_giant_2017}
Kim D, Bowman C, {Del Bonis-O'Donnell} J~T, Matzavinos A and Stein D 2017 {\em
  Phys. Rev. Lett.\/} {\bf 118} 048002

\bibitem{mahendran_protein_2012}
Mahendran K~R, {Romero-Ruiz} M, Schl{\"o}singer A, Winterhalter M and
  Nussberger S 2012 {\em Biophysical Journal\/} {\bf 102} 39--47

\bibitem{hepp_kinetics_2016}
Hepp C and Maier B 2016 {\em Proc. Natl. Acad. Sci. USA\/} {\bf 113}
  12467--12472

\bibitem{uhl_force-dependent_2018a}
Uhl M and Seifert U 2018 {\em Phys. Rev. E\/} {\bf 98} 022402

\bibitem{manrubia_stochastic_1999}
Manrubia S~C and Zanette D~H 1999 {\em Phys. Rev. E\/} {\bf 59} 4945--4948

\bibitem{evans_diffusion_2011}
Evans M~R and Majumdar S~N 2011 {\em Phys. Rev. Lett.\/} {\bf 106} 160601

\bibitem{meylahn_large_2015}
Meylahn J~M, Sabhapandit S and Touchette H 2015 {\em Phys. Rev. E\/} {\bf 92}
  062148

\bibitem{fuchs_stochastic_2016}
Fuchs J, Goldt S and Seifert U 2016 {\em Europhys. Lett.\/} {\bf 113} 60009

\bibitem{jackson_classical_1999}
Jackson J~D 1999 {\em Classical electrodynamics\/} 3rd ed ({New York}: {Wiley})
  ISBN 978-0-471-30932-1

\bibitem{Note1}
\url {https://github.com/muhl/StepPotentialPropagator}

\bibitem{simon_what_1992}
Simon S~M, Peskin C~S and Oster G~F 1992 {\em Proc. Natl. Acad. Sci. USA\/}
  {\bf 89} 3770--3774

\bibitem{peskin_cellular_1993}
Peskin C~S, Odell G~M and Oster G~F 1993 {\em Biophys. J.\/} {\bf 65} 316--324

\end{thebibliography}

\end{document}